\documentstyle[preprint,aps,prl]{revtex}  
\begin{document}

\title{Solution of two channel spin-flavor Kondo model}
\author{Jinwu  Ye}
\address{
Physics Laboratories,       
Harvard  University, Cambridge, MA, 02138}
\date{\today}
\maketitle
\begin{abstract}
  We investigate a model where an impurity couples to both the spin and
  the flavor currents of the two channel conduction electrons.  This model
  can be used as a prototype model of a {\em magnetic} impurity tunneling
  between two sites in a metal and of some heavy fermion systems where the
  ground state of the impurity has a fourfold degeneracy.  The system is 
  shown to flow to a doubly degenerate {\em non fermi-liquid}(NFL) fixed point;
  the thermodynamic quantities show NFL behaviors, but the transport quantities
  show fermi liquid (FL) behaviors . A spin-flavor coupling double tensor term
  is shown to drive the system to one of the two singlet FL fixed points. The
  relation with $ SU(4) $ Coqblin-Schrieffer model is studied.  The implications
  on the possible experiments are given.
\end{abstract}
\narrowtext

  Vlad\'{a}r and Zawadowski \cite{zaw}
  suggested that the model of a non-magnetic heavy particle tunneling
  between two sites in a metal can be mapped to the two channel Kondo model
  in which  the roles of channels
  and spins in the original formulation are interchanged;
  this mechanism has been proposed to explain the
  non-fermi liquid (NFL) behavior observed in the conductance
  signals of the ballistic metal contacts and in some heavy fermion systems
  \cite{ralph}.
  Recently, Ref.\cite{fisher,hopping} reinvestigated
   this mapping in details.
  If the heavy particle also carries spin, for example,
  a muon is injected into a metal, then the heavy particle also interacts
  with conduction electron gas in the channel(the real spin)
  sector. Simply speaking, the model of
  a {\em magnetic} heavy particle tunneling between two sites in a metal
  can be mapped to the two channel spin-flavor Kondo model (2CSFK)
  where the impurity, having both the spin and the flavor degree of freedoms,
  couples to both the spin and the flavor currents of the two channel
  conduction electrons.
  D. L. Cox \cite{cox} pointed out that when
  the ground state of the impurity is a orbital (non-magnetic) doublet,
  the resulting two channel quadrupolar Kondo effects may explain the
  NFL behaviors observed in some heavy fermion systems.
  However, under some symmetry conditions, it is also possible that
  the impurity has a four fold degenerate ground state which can be
  described by quadrupole (non-magnetic) moment
  and magnetic moment (channel) \cite{cox,pang}, the resulting
  2CSFK maybe a appropriate model to describe the low temperature
  physics of such systems. 

  Various kinds of methods like Bethe ansatz \cite{bethe},
  Conformal field theory (CFT) \cite{affleck1},
  Numerical Renormalization Group(NRG) \cite{affleck2},
  Abelian bosonization \cite{emery} etc. can be
  used to solve this model.
  Using NRG, Pang \cite{pang} investigated the 2CSFK
  ( both with and without double tensor term Eq.\ref{doubletensor}),
  unfortunately, most of his results are incorrect.
  Based on the pioneering work of Affleck and Ludwig (AL) \cite{affleck1},
  Emery and Kivelson (EK) \cite{emery}, Maldacena and Ludwig (ML) \cite{ludwig},
  the author developed a simple and powerful method to study certain class of
   quantum impurity models \cite{powerful}. This method  circumvent the difficulty
  to identify the fusion rules of CFT approach.
  In this paper, using the method, we reinvestigate this model.
   We find that the system flows to a doubly
  degenerate ( $ g=2 $ )  NFL  fixed point where the conduction electrons
  suffer phase shift $ \pm \pi/4 $,
   the spin-spin (also flavor-flavor) correlation,
  the pairing correlation and the transport properties of the
  conduction electrons show {\em FL } behaviors.
  However, because the leading irrelevant operators are {\em NFL}
  operators, the thermodynamic quantities show {\em NFL}
  behaviors.  A spin-flavor coupling double tensor
  term Eq.\ref{doubletensor} is shown to break the doublet and
  drive the system to the singlet ( $ g=1 $ ) fermi liquid
  fixed points with phase shifts either $ \pi/4 $ or $ -\pi/4 $.
   All the irrelevant operators
  near these two singlet FL fixed points are shown to be
  FL operators.
   The relation between 2CSFK and
   $ SU(4) $ Coqblin-Schrieffer model is also studied.
  The important lesson we learned from this model is that it is possible
  that the transport properties show {\em FL} behaviors,
   but thermodynamic quantities
  show {\em NFL} behaviors.

   We consider the following Hamiltonian of the 2CSFK
 \begin{eqnarray}
 H &= & i v_{F} \int^{\infty}_{-\infty} dx 
   \psi^{\dagger}_{i \alpha }(x) \frac{d \psi_{i \alpha }(x)}{dx}
   + \sum_{a=x,y,z} \lambda^{a}_{s} J^{a}_{s}(0)  S^{a} 
   + \sum_{a=x,y,z} \lambda^{a}_{f} J^{a}_{f}(0)  F^{a}  \nonumber \\
  & + & h_{s} ( \int dx J^{z}_{s}(x) + S^{z} )
   +h_{f} ( \int dx J^{z}_{f}(x) + F^{z} )
\label{kondof}
 \end{eqnarray}
   where $ J^{a}_{s}(x) = \frac{1}{2} \psi^{\dagger}_{i \alpha }(x)
  ( \sigma^{a} )_{\alpha \beta} \psi_{i \beta }(x) $
   and $ J^{a}_{f}(x) =\frac{1}{2} \psi^{\dagger}_{i \alpha }(x)
  ( \sigma^{a} )_{i, j } \psi_{j \alpha }(x) $
   are the spin and the flavor currents of conduction electrons respectively.
   $ S^{a} , F^{a} $ $  (a=x,y,z) $  are the impurity spin and flavor
    respectively, the impurity has $ g= 2 \times 2 =4 $ states.
   $ h_{s}, h_{f} $ are the uniform magnetic and flavor fields respectively.
   In most of this paper, we 
  limit our discussions 
   to the  $ O(2) $ symmetry in both spin and flavor sectors.

   If $\lambda_{s}^{a}=\lambda_{s}, \lambda_{f}^{a}=\lambda_{f},
   h_{s}=h_{f}=0 $, then the Hamiltonian Eq.\ref{kondof}
   has global $ SU_{s}(2) \times SU_{f}(2) \times U_{c}(1) $  symmetry.
  If $ \lambda_{s}=\lambda_{f}$, there is also a exchange
   ( $Z_{2} $ ) symmetry between spin and flavor sectors.
  The model also enjoys Particle-Hole (P-H ) symmetry :
  $\psi_{i \alpha}(x) \rightarrow \epsilon_{i j} \epsilon_{\alpha \beta}
    \psi^{\dagger}_{j \beta}(x) $ and Time-Reversal (T)
   symmetry :
  $\psi_{i \alpha}(x) \rightarrow
   (\sigma^{2})_{i j} (\sigma^{2})_{\alpha \beta}
    \psi_{j \beta}(-x), \vec{ S} \rightarrow - \vec{ S}, \vec{F }
    \rightarrow -\vec{F} $ .

       In the following, we closely follow the notations of Emery-Kivelson
  \cite{emery}.  Following the three standard steps in EK solution:
  {\em step 1} : write Hamiltonian in terms of the chiral bosons
  introduced in \cite{emery}. {\em step 2}:
  make the  canonical transformation $ U= \exp [ i S^{z} \Phi_{s}(0)
    + i F^{z} \Phi_{f}(0) ] $. {\em step 3}: make the
    following fermionization:
\begin{eqnarray}
S^{x} &= & \frac{ \widehat{a}_{s}}{\sqrt{2}},~~~
S^{y}= \frac{ \widehat{b}_{s}}{\sqrt{2}},~~~
S^{z}= -i \widehat{a}_{s} \widehat{b}_{s} =d^{\dagger}_{s} d_{s}-\frac{1}{2}     \nonumber \\
F^{x} &= & P_{F} \frac{ \widehat{a}_{f}}{\sqrt{2}},~~~
F^{y}= P_{F} \frac{ \widehat{b}_{f}}{\sqrt{2}},~~~
F^{z}= -i \widehat{a}_{f} \widehat{b}_{f}=d^{\dagger}_{f}d_{f}-\frac{1}{2}        \nonumber \\
 \psi_{sf,i} & = & \frac{1}{\sqrt{2}}( a_{sf,i} - i b_{sf,i} ) =
  \frac{P_{sf}}{\sqrt{ 2 \pi a}} e^{-i \Phi_{sf} }     \nonumber   \\
 \psi_{s,i} & = & \frac{1}{\sqrt{2}}( a_{s,i} - i b_{s,i} )=
 \frac{P_{s}}{\sqrt{ 2 \pi a}} e^{-i \Phi_{s} }       \nonumber   \\
 \psi_{f,i} & = & \frac{1}{\sqrt{2}}( a_{f,i} - i b_{f,i} )=
 \frac{P_{f}}{\sqrt{ 2 \pi a}} e^{-i \Phi_{f} }   
\end{eqnarray}
  Where the cocyle factors are $ P_{F}=e^{i \pi (d^{\dagger}_{s}d_{s}+ N_{sf})},
  P_{sf}=e^{i \pi d^{\dagger}_{s} d_{s}},
  P_{s}= e^{i \pi (d^{\dagger}_{s}d_{s}+ d^{\dagger}_{f} d_{f} +N_{sf})},
  P_{f}= e^{i \pi (d^{\dagger}_{s}d_{s}+ d^{\dagger}_{f} d_{f} +N_{sf}+N_{s})} $.

    The transformed Hamiltonian 
  $ H^{\prime}= H_{sf} + H_{s} +H_{f} + \delta H $ can be written
  in terms of Majorana fermions as \cite{atten}: 
\begin{eqnarray}
 H_{sf} &= & H_{0sf} + +i \frac{ \lambda_{s} }{\sqrt{ 2 \pi a}} \widehat{b}_{s} a_{sf}(0)
   +i \frac{ \lambda_{f} P_{c} }{\sqrt{ 2 \pi a}}   \widehat{a}_{f} b_{sf}(0)
                                                 \nonumber \\
 H_{s} & = & H_{0s} -i h_{s} \int dx a_{s}(x) b_{s}(x)   \nonumber  \\
 H_{f} &= & H_{0f} -i h_{f} \int dx a_{f}(x) b_{f}(x)   \nonumber  \\
  \delta H &= & O_{s}+ O_{f}, ~~~O_{s}= -\lambda^{z \prime}_{s} \widehat{a}_{s} 
\widehat{ b}_{s} a_{s}(0) b_{s}(0)
  ,~~~ O_{f}= -\lambda^{z \prime}_{f} \widehat{a}_{f}
   \widehat{ b}_{f} a_{f}(0) b_{f}(0)
\label{sfkm}
\end{eqnarray}
  where $ P_{c}=e^{i \pi N_{c}}, \lambda^{z \prime}_{s} = \lambda^{z}_{s} - 2 \pi v_{F}, 
  \lambda^{z \prime}_{f} = \lambda^{z}_{f} - 2 \pi v_{F} $ \cite{pc}. 

From the above Majorana fermion  basis, it is easy to see that
  {\em half} of the impurity spin couple to half of
  the spin-flavor electrons, {\em half} of the impurity flavor couple
  to {\em another} half of the spin-flavor electrons, this fact make the
  stable fixed point of the 2CSFK rather different from that of the 2CK. 

   It is easy to see at $ \lambda^{z}_{s}=\lambda^{z}_{f}=2 \pi v_{F} $,
  the spin and flavor bosons $\Phi_{s}, \Phi_{f} $ completely decouple
  from the impurity, therefore $ \chi^{s}_{imp}=\chi^{f}_{imp}=0 $.
   Because the canonical transformation is a boundary condition
   changing operator \cite{powerful}, this fact leads to the boundary conditions
   on the solvable line
\begin{equation}
 a^{s}_{L}(0)=-a^{s}_{R}(0) ,~~ b^{s}_{L}(0)=-b^{s}_{R}(0) ; ~~~~~
 a^{f}_{L}(0)=-a^{f}_{R}(0) ,~~ b^{f}_{L}(0)=-b^{f}_{R}(0)
\label{bound1}
\end{equation}

  In order to identify the fixed point {\em along} the EK solvable line 
   $ \lambda^{z \prime}_{s} = 0, \lambda^{z \prime}_{f} = 0 $,
  we write $ H_{sf} $ in the Lagrangian form:
\begin{eqnarray}
 S_{sf} &= & S_{0sf} +  \frac{\gamma_{s}}{2} \int d\tau  \widehat{b}_{s}(\tau)
   \frac{ \partial \widehat{b}_{s}(\tau)}{\partial \tau}
   +i \frac{ \lambda_{s} }{\sqrt{ 2 \pi a}} \int d\tau \widehat{b}_{s}(\tau)
   a_{sf}(0,\tau)                   \nonumber \\
 & + & \frac{\gamma_{f}}{2} \int d\tau \widehat{ a}_{f}(\tau)
     \frac{ \partial \widehat{a}_{f}(\tau)}{\partial \tau}
   +i \frac{ \lambda_{f} P_{c} }{\sqrt{ 2 \pi a}} \int d \tau
     \widehat{a}_{f}(\tau)b_{sf}(0,\tau)
\label{rg}
\end{eqnarray}
 
   Following Ref.\cite{powerful}, when performing the RG analysis of Eq.\ref{rg},
 we keep $\lambda_{s} $ and $ \lambda_{f} $ fixed, simple power countings lead to the following
 RG flow equations 
\begin{eqnarray}
\frac{d \gamma_{s}}{d l} = -\gamma_{s}      \nonumber  \\
\frac{d \gamma_{f}}{d l} = -\gamma_{f}     
\end{eqnarray}

  It is easy to see that the {\em only} stable fixed point is located
 at $ \gamma_{s}=\gamma_{f}=0 $ where $ \widehat{b}_{s} $ and $ \widehat{a}_{f} $
  lose their kinetic energies and become two
  Grassmann Lagrangian multipliers;  integrating them out leads to
    the {\em boundary conditions} \cite{powerful} :
\begin{equation}
  a^{sf}_{L}(0)=-a^{sf}_{R}(0),~~~   b^{sf}_{L}(0)=-b^{sf}_{R}(0)
\label{bound2}
\end{equation}

    Boundary conditions \ref{bound1} and \ref{bound2} 
  can be expressed in terms of bosons:
\begin{equation}
 \Phi^{s}_{L}= \Phi^{s}_{R}+ \pi,~~ \Phi^{sf}_{L}= \Phi^{sf}_{R}+ \pi,~~
 \Phi^{f}_{L}= \Phi^{f}_{R}+ \pi
\label{bose1}
\end{equation}
   Assuming no potential scattering, we also have:
\begin{equation}
   \Phi^{c}_{L}=\Phi^{c}_{R} 
\label{bose2}
\end{equation}

   Substituting the above two Eqs. into their standard definitions
  \cite{emery},
  we find all the four Dirac fermions 
 suffer the same phase shift $ \delta= \pm \frac{\pi}{4} $.
 it is well known that in the presence of P-H symmetry, the phase shift
 can only be 0 or $ \pi/2 $, however the ground state of this model is a
 {\em doublet}, one causes phase shift $ \frac{\pi}{4} $, another causes
 phase shift $ -\frac{\pi}{4} $, each breaks P-H symmetry and related
 to each other by P-H transformation \cite{chance}.
  The symmetry of this fixed point is  $ O(6) \times O(2) \sim
   SU(4) \times O(2) $. The finite size $ - l< x <l $ spectrum
  will be that of 4 free Dirac fermions with $ \pm \frac{\pi}{4}$
 phase shift.
 
   The above results can also be reached from CFT. In order to get the 
 finite size spectrum at the stable fixed point, we have to do two fusions
  simultaneously from the free electrons spectrum, one in the spin sector,
  another in the 
 flavor sector. It can be shown explicitly that the resulting spectrum 
 is indeed equivalent to the {\em superposition} of phase shift $ \frac{\pi}{4} $ and
  phase shift $-\frac{\pi}{4} $. As shown in Ref.\cite{powerful}, it is
 more convenient and more compact to calculate the finite size spectrum
 according to the {\em maximum } symmetry $ O(6) \times O(2) $ of the fixed point.
 we find the ground state energy is $\frac{1}{8} $ with the degeneracy 2 \cite{chance}.
 The excitation spectrums are equally spaced with
  $ \Delta E= 1/4 $. The complete finite size spectrum
  of this fixed point is listed in \cite{hopping}.
  However, Pang got {\em a line of NFL fixed points} with continuously changing
  phase shifts \cite{pang,chance}.

  Exactly at this fixed point, the impurity degeneracy is
  $ g= \sqrt{2} \times \sqrt{2} =2 $, so $ S_{imp} ( T=0 ) = \log 2 $.
   One particle S matrix is $ S_{1}= \pm i $, as in 2CK, the
   residual resistivity is also
   half of the unitarity limit $ \rho(0) =\frac{\rho_{u}}{2} $.
 The scaling dimensions of the various fields at this fixed point
 are: $ [ \widehat{b}_{s}]= [ \widehat{a}_{f}]=1/2, [a_{sf}]=[b_{sf}]=3/2 ,
 [ \widehat{a}_{s}]= [ \widehat{b}_{f}]=0 $.
 The impurity spin and flavor correlation functions
 show the typical 2CK {\em NFL} behaviors
  $ \langle S^{z}(\tau) S^{z}(0) \rangle \sim 1/ \tau,
   \langle F^{z}(\tau) F^{z}(0) \rangle \sim 1/ \tau $.
  We find two {\em leading
 irrelevant } operators $ O_{s}, O_{f} $ with dimension 3/2 which will
  generate two dimension 2 operators 
  $ a_{s}(0,\tau) \frac{ \partial a_{s}(0,\tau)}{\partial \tau} + b_{s}(0,\tau) \frac{
   \partial b_{s}(0,\tau)}{\partial \tau},
   a_{f}(0,\tau) \frac{ \partial a_{f}(0,\tau)}{\partial \tau} + b_{f}(0,\tau) \frac{
   \partial b_{f}(0,\tau)}{\partial \tau} $, the other two dimension 2 operators are
   the $ \gamma_{s} $ and $ \gamma_{f} $ terms in Eq.\ref{rg}.
 In all, there are
  four {\em sub-leading irrelevant } operators with dimension 2.
   The straightforward extension of the
  CFT analysis in Ref. \cite{line} shows that there are
   two dimension 3/2 operators $ T_{00}^{s},
  T_{00}^{f} $ and four dimension 2 operators $ P_{1}^{s}, P_{2}^{s};
   P_{1}^{f}, P_{2}^{f}$, in consistent with the above analysis.

  Although the ground state is a doublet which are spin singlet, 
  flavor singlet with charges $ \pm 1 $,
  however, all {\em allowed} boundary operators must be {\em neutral}, therefore, will
  {\em not} mix the doublet \cite{ferro}; the {\em non-degenerate} perturbation
  theory can still be used to calculate the thermodynamic and transport
  properties.  The only operator which can mix this doublet is the spin-singlet,
  flavor-singlet pair operator, although this operator is not a {\em allowed}
  boundary operator, but it is responsible for the enhanced pairing susceptibility
  of the 2CK \cite{powerful}. In the case of FL boundary conditions as happened in the present
  model, this operator vanishes \cite{resitivity,powerful}. There is {\em no} pairing susceptibility
  enhancement in the 2CSFK.
  The electron spin-spin (flavor-flavor) correlation and the
  pairing operator correlation are
  exactly the {\em same} with those of the free electrons.
  Following the calculations in Refs.\cite{powerful,flavor}, 
  it is easy to see that
  the low temperature correction to the electron resistivity is 
 controlled by the {\em sub-leading} irrelevant operators with dimension 2
 in contrast to 2CK where it is controlled by the {\em leading}
 irrelevant operators with dimension 3/2.  At low temperature,
 the second order perturbation in these
 operators yield $ \rho (T) - \frac{ \rho_{u}}{2} \sim  T^{2} $.
 However, similar to the 2CK, the impurity
 susceptibility and specific heat are still controlled by the
  dimension 3/2 leading irrelevant operators.
  $ C_{imp} \sim ( \lambda^{ \prime 2}_{zs} + \lambda^{ \prime 2}_{zf} )
  T \log T, \chi^{s}_{imp} \sim \lambda^{ \prime 2}_{zs} \log T,
  \chi^{f}_{imp} \sim \lambda^{ \prime 2}_{zf} \log T $, the {\em combined}
  Wilson ratio $ R= \frac{ T( \chi^{s}_{imp} + \chi^{f}_{imp} )}{
  C_{imp}} $ is universal \cite{bhatt}.  Pang got totally different results for the physical
  measurable quantities mentioned above.

   The potential scattering term can be added to the Hamiltonian \ref{kondof},
 it breaks P-H symmetry and causes a continuous phase shift
  in the charge sector, the symmetry of the fixed line is
  $ SU(4) \times U(1) \sim U(4) $ with $ g=2 $.

 A {\em spin-flavor coupling} double tensor term can also be added:
\begin{equation}
  \lambda^{ab} J^{a b}(0) S^{a} F^{b}
\label{doubletensor}
\end{equation}
   where $ J^{ab}(x) = \frac{1}{4}
            \psi^{\dagger}_{i \alpha}(x) ( \sigma^{a})_{i j }
            (\sigma^{b})_{\alpha \beta} \psi_{j \beta}(x) $.

  It can be shown that the double tensor term still respects
   global $ SU_{s}(2) \times SU_{f}(2) \times U_{c}(1) $  symmetry if
 $ \lambda^{ab} =\lambda $, it
  is {\em odd } under P-H, 
  {\em even} under T. For simplicity, we consider $ O_{s}(2) 
   \times O_{f}(2) \times U_{c}(1) $
   case where $ \lambda_{11}= \lambda_{12}= \lambda_{21}= \lambda_{22},
  \lambda_{13}= \lambda_{23}, \lambda_{31}= \lambda_{32} $.

   It is instructive to compare the model with the $ SU(N) $
   Coqblin-Schrieffer(CS) model \cite{read}
\begin{equation}
  H_{cs}=  \int \frac{d k}{2 \pi} C^{\dagger}_{k \alpha} C_{k \alpha}
        + J [ C^{\dagger}_{\alpha}(0) C_{\beta}(0)
          d^{\dagger}_{\beta} d_{\alpha}-\frac{1}{N}C^{\dagger}_{\alpha}(0)
           d^{\dagger}_{\beta} d_{\beta} C_{\alpha}(0)]
\label{cs}
\end{equation}
   where $\alpha, \beta=1,\cdots,N $, with the constraint
    $ d^{\dagger}_{\beta} d_{\beta}=m $, the summation over repeated
   indices is implied.

 The phase shift at the FL fixed point of the $ SU(N) $
  CS model satisfies
 the Friedel sum rule $ \delta= \frac{m}{N} \pi $.
  Under P-H transformation $ d_{\alpha}
 \rightarrow d^{\dagger}_{\alpha},
  C_{k,\alpha} \rightarrow C^{\dagger}_{-k,\alpha} $; $ m \rightarrow N-m $.
  For {\em even} $ N $, if $ m= \frac{N}{2} $, then $ \delta= \frac{\pi}{2} $, the CS
  model has P-H symmetry.

   In contrast to the claim by Pang \cite{pang}, the model described by
  Eqs. \ref{kondof} and \ref{doubletensor} is {\em not}
  equivalent to $SU(4)$ Coqblin-Schrieffer(CS) Model for any values of
 parameters, because the two models have {\em different} global symmetry.

 Following the three standard steps in EK solution,
 the double tensor term Eq.\ref{doubletensor} can be written as:
\begin{equation}
 i \frac{ \lambda_{11} P_{c}}{ 4 \pi a}  \widehat{a}_{s} \widehat{b}_{f}
+ \frac{ \lambda_{13} P_{c}}{ 2 \sqrt{2 \pi a } }
  \widehat{a}_{s} \widehat{b}_{f} \widehat{b}_{s} a_{sf}(0)
-\frac{\lambda_{31}}{\sqrt{ 2 \pi a}}          
 \widehat{a}_{s} \widehat{b}_{f} \widehat{a}_{f} b_{sf}(0)
- i \frac{\lambda_{33} }{2} 
 \widehat{a}_{s} \widehat{b}_{f} \widehat{b}_{s} a_{sf}(0)
 \widehat{a}_{f} b_{sf}(0)
\label{double}
\end{equation}

  The first term in Eq.\ref{double} has scaling
  dimension 0, therefore is a relevant
 perturbation to the fixed point of the 2CSFK. In order to locate the stable
  fixed point, we define a {\em artificial} impurity
 spin $ L^{z}= -i \widehat{a}_{s} \widehat{b}_{f}$, therefore we can
 write Eq.\ref{double} as
\begin{equation}
  L^{z}[ -\frac{ \lambda_{11} P_{c}}{ 4 \pi a} 
 +i \frac{ \lambda_{13} P_{c}}{ 2 \sqrt{2 \pi a }} \widehat{b}_{s} a_{sf}(0)
 -i \frac{\lambda_{31}}{\sqrt{ 2 \pi a}}\widehat{a}_{f} b_{sf}(0)
  + \frac{\lambda_{33} }{2} \widehat{b}_{s} a_{sf}(0)
     \widehat{a}_{f} b_{sf}(0) ]
\label{exact}
\end{equation}

 Under P-H, $ L^{z} \rightarrow -L^{z} $,
  $ L^{z} = \pm \frac{1}{2} $ are {\em exact} 
 eigenstates of Eq.\ref{exact}. It is easy to see $ \lambda_{11} $
 plays the role of the {\em artificial} magnetic field
  which couples to the {\em artificial}
 impurity spin $ L^{z} $. Depending on the sign of $\lambda_{11}$, the 
 impurity ground state will be either $ L^{z} =\frac{1}{2} $ or
 $ L^{z} =-\frac{1}{2} $, the excited state can be projected out without
  affecting the low energy properties,
  the system flows to one of the two $ g=1 $
  stable fixed points where $ \langle \widehat{a}_{s}(\tau)
     \widehat{a}_{s}(0)\rangle
   \sim ( \frac{\lambda^{z \prime}_{s}}{\lambda_{11}})^{2}\frac{1}{\tau},
   \langle \widehat{b}_{f}(\tau) \widehat{b}_{f}(0) \rangle
   \sim ( \frac{\lambda^{z \prime}_{f}}{\lambda_{11}})^{2}\frac{1}{\tau} $, therefore
 $ [ \widehat{a}_{s}]= [\widehat{b}_{f}] = \frac{1}{2} $.
  $ \lambda_{13} $ and $ \lambda_{31} $ in Eq.\ref{exact} can be combined with
  $ \lambda_{s} $ and $\lambda_{f} $ in Eq.\ref{sfkm} respectively, therefore
 the boundary conditions of the two $ g=1 $ fixed points are
 {\em still given} by Eqs.\ref{bose1},\ref{bose2},
  namely both have $ O(4) \times O(2) $ symmetry
  and are P-H conjugated; both have the same finite size spectrum which can be
 achieved by reducing the degeneracies of all the energy levels by 2 of that of the $ g=2 $ fixed point.
 this is another non-trivial
  example of AL's g-theorem \cite{gtheorem}.
 The impurity spin and flavor correlation functions
 show typical FL behaviors
  $ \langle S^{z}(\tau) S^{z}(0) \rangle \sim \langle F^{z}(\tau) F^{z}(0) \rangle \sim 
   \langle S^{z}(\tau) F^{z}(0) \rangle \sim 1/\tau^{2} $.
 The fixed point with $ L^{z}= \frac{1}{2},
 \delta= \frac{\pi}{4}, S=i $ is the {\em same} fixed point
 as $ SU(4) $ CS model with $ m=1 $ local fermion, another with
 $ L^{z}=-\frac{1}{2}, \delta= -\frac{\pi}{4}, S=-i $ is
 the {\em same } fixed point as $ SU(4) $ CS model with
 $ m=3 $ local fermions, therefore both have the {\em same } boundary
  operator contents
 as their corresponding CS models. However, because the two models
 have different global symmetry, the {\em allowed } boundary operators are
 {\em different}.
 
 There are 6 {\em leading irrelevant } operators with scaling
 dimension 2.  $ O_{s} \sim \partial \Phi_{sf} \partial \Phi_{s},
   O_{f} \sim \partial \Phi_{sf} \partial \Phi_{f} $ are two dimension 2 operators,
  The other 4 dimension 2 operators are the same as
  the {\em sub-leading} irrelevant operators near the $ g=2 $ fixed point, but they
  become the {\em leading} irrelevant operators near the $ g=1 $ fixed points.
  The $\lambda_{33}$ term in 
  Eq.\ref{exact} has scaling dimension 4. It is easy to see
  that all the boundary operators must have {\em integer} scaling
  dimensions $ h \geq 2 $, therefore are
  {\em FL} boundary operators.

 Because the P-H symmetry is always absent, the double tensor
 term is always present, this make the experimental observation of
    the {\em mixed} behaviors: $ S_{imp}= \log 2, C_{imp} \sim T \log T,
    \chi_{imp} \sim \log T, \rho(T) \sim \rho(0) + T^{2}$ very unlikely.

%

We thank D. S. Fisher, B. Halperin, I. Smolyarenko especially D. Cox and
 N. Read for helpful discussions.
This research was supported by NSF Grants Nos. DMR 9106237 and DMR9400396.


\begin{references}
\bibitem{zaw} K. Vlad\'{a}r and A. Zawadowski, Phys. Rev. B28, 1564(1983);
              28, 1582(1983); 28, 1596(1983).
\bibitem{ralph} D. C. Ralph and R. A. Buhrman, Phy. Rev. Lett. 69, 2118 (1992);
   D. C. Ralph, A. W. W. Ludwig, Jan von Delft and R. A. Buhrman, {\sl ibid.}
   72, 1064(1994); C. L. Seaman, {\sl et. al}, {\sl ibid.} 67, 2882(1991).
\bibitem{fisher} A. L. Moustakas and D. S. Fisher, Phys. Rev. B 51,6908 (1995),
   53, 4300 (1996) 
\bibitem{hopping} Jinwu Ye, cond-mat/9609076.
\bibitem{cox} D. L. Cox, Phy. Rev. Lett. 59, 1240 (1987)
\bibitem{pang} H. Pang, Phy. Rev. Lett. 73, 2736 (1994)
\bibitem{bethe} N. Andrei and C. Destri, Phys. Rev. Lett. 52, 364(1984);
    A. M. Tsvelik and P. B. Wiegmann, Z. Phys. B54, 201 (1984); A. M. Tsvelik,
    J. Phys. C18, 159(1985). 
\bibitem{affleck1} I. Affleck and A. W. W. Ludwig, Nucl. Phys. B360, 641(1990) 
\bibitem{affleck2} I. Affleck, A. W. W. Ludwig, H. B. Pang and D. L. Cox
 Phys. Rev. B45, 7918(1992) 
\bibitem{emery} V. J. Emery and S. Kivelson, Phys. Rev. B46, 10812(1992)
  , Phys. Rev. Lett. 71 (1993) 3701.
\bibitem{ludwig} J. M. Maldacena and A. W. W. Ludwig,
                 to be published in Nucl. Phys. B.
\bibitem{line} Jinwu Ye, Phys. Rev. Lett. 77, 3224 (1996).
\bibitem{powerful} Jinwu Ye, cond-mat/9612029 and references therein.
\bibitem{chance} This doublet is obtained by taking $ l \rightarrow \infty $ first,
   then $ T \rightarrow 0 $.
   The ground state degeneracy of finite size spectrum is obtained by
   $ T \rightarrow 0 $ first, then $ l \rightarrow \infty $, it 
   turns out also to be 2 with charge $ \pm 1 $. 
   These two degeneracies are usually {\em different} because the two limits
   donn't commute \cite{affleck1}.
   Pang's NRG results can only give the {\em finite size} degeneracy.
\bibitem{flavor} Jinwu Ye, cond-mat/9609057.
\bibitem{atten} The differences between $\psi_{s,i},\psi_{sf,i} $
  and $ \psi_{s}, \psi_{sf} $  are neglected in the following without affacting
  any results.
\bibitem{pc} $ P_{c} $ is simply a constant, because $ N_{c} $ is conserved.
\bibitem{ferro} For ferromagnetic Kondo model, the ground state degeneracy is also 2,
 but the two states are {\em mixed} by the irrelevant operator which is the spin fliping part of
 the Kondo interaction. This is responsible for the NFL behaviours in the one
 channel ferromagnetic Kondo model.
\bibitem{resitivity} A. W. W. Ludwig and I. Affleck, Nucl. Phys. B428, 545(1994); 
 I. Affleck and A. W. W. Ludwig, Phys. Rev. B48,7297(1993). 
\bibitem{bhatt} N. B. Bhatt and D. S. Fisher, Phys. Rev. Lett. 68, 3072 (1992).
   They discussed the metallic phase of disordered metal and showed
 that the low temperature thermodynamic quantities show NFL behaviours, but
 the charge transport still shows FL behaviours. This is due to
  {\em residual} local moments which are essentially decoupled from the
   conduction electron fluids at any low temperature.
\bibitem{gtheorem} I. Affleck and A. W. W. Ludwig,
  Phy. Rev. Lett. 67, 161(1991)
\bibitem{read} N. Read and D. M. Newns, J. Phys. C16, 3273(1983)
\end{references}
\end{document}